# The unblinking eye on the sky

Eric C. Bellm and Shrinivas R. Kulkarni

From near-Earth asteroids to superluminous supernovae and counterparts to gravitational wave sources, the Zwicky Transient Facility will soon scan the night sky for transient phenomena, explain Eric Bellm and Shrinivas Kulkarni.

The Zwicky Transient Facility (ZTF) is a next-generation optical synoptic survey motivated by the search for rare and fast-evolving optical transients. ZTF succeeds the Palomar Transient Factory (PTF) survey [1, 2] and the intermediate Palomar Transient Factory (iPTF) survey on the 48-inch Samuel Oschin Schmidt telescope (P48) at Palomar Observatory. PTF began operations in 2009 with large area surveys to characterize the optical transient sky. The iPTF has operated for the past four years, emphasizing same night multi-colour follow-up of transient candidates, most recently using a novel ultra-low resolution robotic spectrograph on the robotic Palomar 60-inch telescope (P60). The survey camera on the P48 (the CFH12K, originally used on the Canada-France-Hawaii telescope) had a field-of-view of eight square degrees.

ZTF's custom mosaic camera [3, 4] will maximize the discovery rate of transients on the P48 [5]. It completely fills the available 47 deg$^2$ focal plane of the P48 with sixteen charge-coupled devices of 6144 x 6160 pixels each and reduces readout and slew overheads to less than 15 seconds per exposure. The resulting order-of-magnitude improvement in survey speed relative to PTF will enable ZTF to survey at high cadence over a wide area. The exposure time of 30 s was deliberately chosen to achieve a moderate depth ($m_r \sim 20.5$; $5\sigma$) so that follow-up could be undertaken by existing telescopes, for example, the P60 for ultra-low resolution spectroscopy, 1-m telescopes for multi-band photometry, and 3-m to 5-m class telescopes for low-resolution spectroscopy.

ZTF will pursue a broad range of time-domain science, ranging from Near-Earth Asteroids to the study of distant superluminous supernovae. The "Celestial Cinematography" survey undertaken in two bands will cover a large fraction of the observable sky over three nights. This survey is expected to detect over five thousand supernovae of type Ia per year (thus helping to improve the precision of the Hubble diagram at low redshift), over a hundred superluminous supernovae, and a dozen tidal disruption events (TDEs). The survey will allow a thorough census of supernovae of all types within 200 Mpc. Typically, the region of the sky least observed by synoptic surveys is the Galactic plane, where most of the stars in our Galaxy are located. To rectify this lacuna, ZTF will sweep the Galactic equator nightly. This Plane survey will find active stars, measure stellar rotation periods, and discover a very large number of variable stars as well as cataclysmic variables, novae, and other accretion-powered stellar systems.

Another higher-cadence survey is designed to study explosions at early times. Early observations of supernovae can place radius and binarity constraints on the supernova progenitor, identify shock interactions, and enable discovery of progenitor emission signatures in the photo-ionized circumstellar wind of core-collapse supernovae through "flash" spectroscopy. Two models have been suggested for the origin of type Ia supernovae: a coalescence of two white dwarfs and the explosion of a white dwarf after it accretes sufficient matter from a companion. For the latter, the presence of declining emission in the UV (observed by the Swift X-ray Observatory) is a certain indication. More intriguingly, ZTF may discover on-axis and off-axis emission from gamma-ray burst afterglows and constrain the proportion of other types of extra-galactic explosions (such as the putative baryon-loaded "dirty fireballs" [6]). ZTF will be reasonably effective in discovering and following up small streaking Near-Earth Asteroids, although it is not optimized to study them (for that see the newly-commissioned ATLAS facility) [7]. Finally, with its fast areal survey speed and rapid response, ZTF can efficiently search for optical counterparts to Advanced LIGO gravitational wave detections.

We are now solidly in the era of "bright" star and bright transient astronomy. For instance, the depth of ZTF and ATLAS are similar to that of ESA's Gaia mission. While Gaia delivers exquisite astrometric and photometric measurements, ZTF provides a rich trove of time-series observations on a variety of timescales. The combination can be expected to be particularly valuable for stellar astronomy. There will be regular data releases of images, catalogues and lightcurves for all ZTF data to enable a wide variety of archival studies. The aforementioned Celestial Cinematography and Plane surveys will issue near real-time transient alerts. Apart from the science returns from the planned surveys, ZTF is expected to provide a valuable platform to develop new methodologies for the burgeoning field of time-domain astronomy.

We expect to achieve first light in July of 2017. The primary program will begin after a six-month commissioning period. In the first year priority will be given to building up of co-added reference images of the sky. At the end of the second year there will be review of lessons learnt and science returns, and the survey parameters will be readjusted accordingly. ZTF is expected to serve as a prototype for the Large Synoptic Survey Telescope (survey start 2022) which has a vastly larger data stream. ZTF is a public–private partnership, with approximately equal support from the National Science Foundation through the Mid-Scale Innovations Program (MSIP) and a consortium of Universities inside and outside the US.

**Author Information**

**Affiliations**

Eric C. Bellm and Shrinivas R. Kulkarni are at the California Institute for Technology, 1200 East California Boulevard, Pasadena, CA 91125, USA.

**Corresponding authors**

Correspondence to Eric C. Bellm (ebellm@caltech.edu) or Shrinivas R. Kulkarni (srk@astro.caltech.edu).

**References**

[1] Law, N. M. *et al*. *Publ. Astron. Soc. Pacific* 121, 1395–1408 (2009).

[2] Rau, A. *et al*. *Publ. Astron. Soc. Pacific* 121, 1334–1351 (2009).

[3] Smith, R. M. *et al*. In *Proc. SPIE* Vol. 9147, 79 (eds Ramsay, S. K. *et al.*) 914779 (SPIE, 2014).

[4] Dekany, R. *et al*. In *Proc. SPIE* Vol. 9908 (eds Evans, J. C. *et al.*) 99085M (SPIE, 2016).

[5] Bellm, E. C. *Publ. Astron. Soc. Pacific* 128, 084501 (2016).

[6] Cenko, S. B. *et al. Astrophys. J.* 769, 130 (2013).

[7] Waszczak, A. *et al*. *Publ. Astron. Soc. Pac.* 129, 034402 (2017).

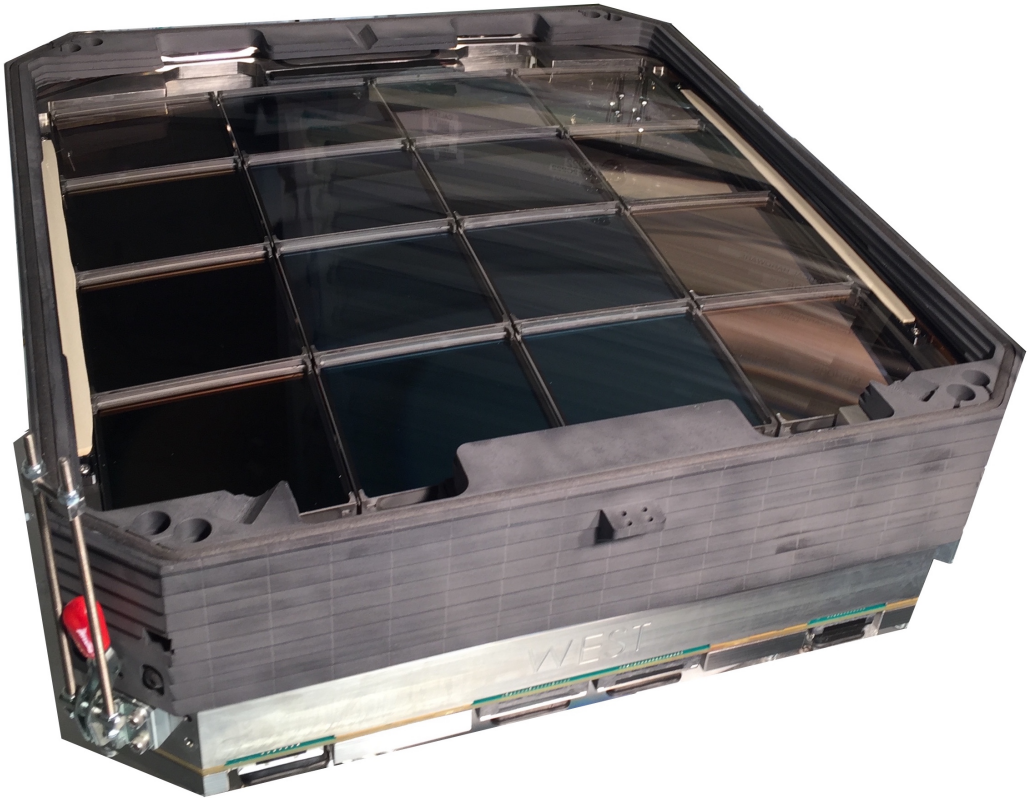

Roger Smith/Michael Feeney, Caltech Optical Observatories

**Figure 1 | The ZTF camera during laboratory testing.** The CCD mosaic is 560mm from corner to corner, which comparable in physical size to the Dark Energy Camera (DECam; 525mm diameter) and the Large Synoptic Survey Telescope (LSST; 640mm diameter). Due to ZTF's coarser pixel scale, however, its field of view of 47 deg$^2$ is much larger than that of DECam (3 deg$^2$) and LSST (~10 deg$^2$).